\documentclass[epjST]{svjour}
\usepackage[numbers, sort, sort&compress]{natbib}
\usepackage{graphics}
\usepackage{dsfont}
\begin{document}
\title{Non-invasive analysis of blood-brain barrier permeability based on wavelet and machine learning approaches}
\author{Nadezhda Semenova\inst{1,2}\thanks{\email{nadya.i.semenova@gmail.com}}
\and Konstantin Segreev\inst{1} \and Andrei Slepnev\inst{1} \and Anastasia Runnova\inst{1,3} \and Maxim Zhuravlev\inst{1,3} \and Inna Blokhina\inst{1} \and Alexander Dubrovsky\inst{1} \and Oxana Semyachkina-Glushkovskaya\inst{1,4} \and J\"{u}rgen Kurths\inst{1,4,5} 
}                     
%
%
\institute{Saratov State University, Astrakhanskaya str., 83, Saratov, 410012, Russia \and D\'{e}partement d'Optique P. M. Duffieux, Institut FEMTO-ST,  Universit\'{e} Bourgogne-Franche-Comt\'{e} CNRS UMR 6174, Besan\c{c}on, France \and State Medical University, B. Kazachaya str., 112, Saratov, 410012, Russia \and Physics Department, Humboldt University, Newtonstrasse 15, 12489 Berlin, Germany \and Potsdam Institute for Climate Impact Research, Telegrafenberg A31, 14473 Potsdam, Germany}
%
%
\abstract{
The blood-brain barrier plays a decisive role in protecting the brain from toxins and pathogens. The ability to analyze the BBB opening (OBBB) is crucial for the treatment of many brain diseases, but it is very difficult to noninvasively monitor OBBB. In this paper we analyse the EEG series of healthy rats in free behaviour and after music-induced OBBB. The research is performed using two completely different methods based on wavelet analysis and machine learning approach. Both methods enable us to recognize OBBB and are in a good agreement with each other. The comparative analysis was carried out using F-measures and ROC-curves.}
%
%
\maketitle
\section{Introduction}
\label{intro}

The blood-brain barrier (BBB) is a highly selective barrier, which is formed by microvascular endothelial cells surrounded by pericytes and perivascular astroglia. It controls the penetration of blood-borne agents into the brain or the release of metabolites and ions from the brain tissue to blood \cite{Abbott2005,Abbott2006,Abbott2010}. Therefore, the BBB plays a vital role in protecting the brain against pathogens and toxins. The BBB disruption is associated with aging, dementia, multiple sclerosis, Alzheimer’s disease, stroke, brain trauma, infection and tumors \cite{Rosenberg2012}. The permeability of BBB can be varied by neuroendocrine regulation and may play a protective role in injury and stroke \cite{Pan1997,Pan1999,Pan2001,Pan2006}. 

The ability to analyze the BBB opening (OBBB) is crucial for the treatment of brain diseases, and it is very difficult to noninvasively monitor OBBB~\cite{Chassidim2013,Heye2014}. Therefore, there is a strong need to create many save methods of the assessment of the BBB permeability in clinical practice. There has been substantial progress over the past few years~\cite{Chassidim2013,Heye2014}. Magnetic resonance tomography (MRI) is an often used technique for the monitoring of OBBB ~\cite{Heye2014}. However, MRI is bulky and cannot be used bedside. The MRI requests the use of contrast agents, which can be even toxic~\cite{Rogosnitzky2016,Kaller2020}. The latter limits its continuous application and usage, especially in the case of children and patients with kidney pathology \cite{Elbeshlawi2018,Perazella2008}. Therefore, the development of novel promising real-time, bedside, non-invasive, label-free, economically beneficial and readily applicable methods is of highest actual importance, and solving this problem would open a novel era in effective diagnosis and therapy of brain diseases that cause acute and chronical OBBB.

In this paper we consider two different approaches for analysing the OBBB by using electroencephalogram (EEG) time series of healthy rats. 
The first method is based on the wavelet analysis of EEG signals. This approach is widely used in many biomedical papers \cite{Unser,Subasi2007}, partially in epilepsy research \cite{Adeli2003,Tzallas2009}, and sleep staging \cite{Bajaj2013,Tripathy2020}, which confirms its applicability in these tasks. Nowadays, the wavelet analysis is one of the main tools for processing data in real time and creating brain-computer interface (BCI) devices~\cite{djamal2020brain,nguyen2015eeg,hettiarachchi2014motor}. 

The second method uses a machine learning approach to analyze such EEG signals. Recently, machine learning has been increasingly used in brain activity analysis \cite{Davatzikos2005,Palaniappan2007,Wei2018,Tian2020}. The application of machine learning to recognize features of brain activity leads to the use of artificial neural networks (ANNs) of different types and topologies. Basically, ANNs do not analyze the EEG signal in its pure form, but use some characteristics based on the EEGs. For example, power spectral density \cite{Ronzhina2012}, statistical features \cite{Hassan2015}, wavelet transform \cite{Hassan2017}, combinations of energies of different frequency bands \cite{Hsu2013} etc. Typically, deep neural networks (DNNs) are used for classification and recognition tasks. As one of examples, the automatic scoring of sleeping stage was recently provided by applying a convolutional neural network \cite{FB2020}. In this work we use DNN with feedforward coupling.

Despite the fact that both methods are based on completely different types of analysis, they show a good ability in recognizing the OBBB. BBB permeability can be evaluated in real time, searching for the EEG signal features, which can be found out by using both wavelet analysis and machine learning approaches.

\section{Experiment design and data recording}
The experiments were conducted on the same five adult male Wistar-Kyoto rats (250-280 g) in awake state and during opened BBB (OBBB). All procedures and experiments were done in accordance with the “Guide for the Care and Use of Laboratory Animals”~\cite{Albus2012}. The experimental protocols were approved by the Local Bioethics Commission of the Humboldt University and the Saratov State University. The animals were kept in a light/dark environment with the lights on from 8:00 to 20:00 and fed ad libitum with standard rodent food and water. The ambient temperature and humidity were maintained at $24.5\pm 0.5 ^{\circ} C$ and 40\%–60\%, respectively. 

Ten days before the experiments, an optical window, a polyethylene catheter, and two epidural EEG electrodes for chronic recordings of electrocorticogram (ECoG) in free behavior were implanted; after surgery, animals were housed individually. The first EEG1-recording was recorded over an hour from 1 to 2 p.m. 10 days after the electrode implantation operation.

The day after the baseline EEG1-recording, the rats were underwent to intermittent music. The detailed description of the protocol of music-induced OBBB is described in~\cite{SGl2020}. To produce the music (70-90-100 dB and 11-10000 Hz, Scorpions “Still loving you”) we used loudspeaker (ranging of sound intensity 0-130 dB, frequencies 63-15000 Hz; 100 V, Yerasov Music Corporation, Saint Petersburg, Russia). The repetitive music exposure was performed using a sequence of: 60 sec – music on and then 60 sec – music off during 2h. The sound level was measured directly in a cage of animals using the sound level meter (Megeon 92130, Russia).
During the 60 min music-off, we performed \textit{in vivo} real time fluorescent microscopy of OBBB for Evans Blue (EB) dye (i.v.), which we detected as bright fluorescence around the cerebral microvessels in awake behaving rats. When we observed the optical signs of EB leakage, EEG recordings in rats were performed during 60 min (EEG2). The choice of time of EEG recordings was related to our previous results discovering the window of music-induced OBBB that is 60-180 min after sound exposure~\cite{semyachkina2017application,zinchenko2019pilot,SGl2020}.
After EEG recordings, the rats decapitated and their brains were collected for confirmation of OBBB for ED using confocal microscopy. The EEG recordings were collected and analyzed using wavelet and machine learning approaches.

A two-channels cortical EEG/one channel electromyogram (EMG) (Pinnacle Technology, Taiwan) were recorded as follows. The rats were implanted two silver electrodes (tip diameter 2–3 {$\mu$}m) located at a depth of 150 {$\mu$}m in the coordinates (L: 2.5 mm and D: 2 mm) from Bregma on either side of the midline under inhalation anesthesia with 2\% isoflurane at 1L/min $N_2O/O_2$ – 70:30. The head plate was mounted and small burr holes were drilled. Afterward, EEG wire leads were inserted into the burr holes on one side of the midline between the skull and the underlying dura. EEG leads were secured with dental acrylic. An EMG lead was inserted in the neck muscle. Ibuprofen (15 mg/kg) for the relief of postoperative pain was provided in their water supply for two to three days prior to surgery and for three or more days post-surgery. The rats were allowed 10 days to recover from surgery prior to beginning the experiment.

All EEG recordings were supplemented with synchronous videography, on the basis of which a neurophysiologist marked out the states of behavioral sleep and wakefulness of animals. Since the standard sleep staging rules for rats are not available currently, we referred the visual scoring criteria from these studies~\cite{Xie2013,Hablitz2019}. Wakefulness, NREM, and REM were visually scored in 10 s epochs. EEG activity was measured and compared in rats in awake state, during sleep, and after OBBB on the same rats. Wakefulness was defined as a desynchronized EEG with low amplitude and high-frequency dynamics ($>$10\%, 8-12 Hz) and relatively high-amplitude EMG. NREM sleep was recognized as synchronized activity with high amplitude, which is dominated by low-frequency delta waves (0–4 Hz) comprising $>$30\% of EEG waveforms/epoch and a lower amplitude EMG. REM was identified by the presence of theta waves (5–10 Hz) comprising $>$20\% of EEG waveforms/epoch with a low EMG amplitude. 

Thus, for each of the animals the EEG signals were recorded before and after music-induced OBBB -- EEG1 and EEG2, respectively.

\section{Recognition of OBBB using time-frequency analysis}
\label{sec:BBB_vawelet}

Today, one of the generally accepted nonlinear methods for processing and detecting of oscillatory activity in biomedical signals is continuous wavelet transformation~(CWT)~\cite{Lachaux:2002_BrainCoherence,feng2017multi,khare2020classification}. CWT is sufficiently resistant to abrupt changes in the frequency composition of the analyzed experimental signals, which makes it possible to adequately analyze rather short time intervals of highly nonstationary signals.
The CWT for an arbitrary signal $x(t)$  in the general form is defined as follows:
\begin{equation}
\label{kor18e}
W(s,t_0)=\int\limits_{-\infty}^{+\infty}x(t)\psi_{s,t_0 }^*(t)dt,
\end{equation}
where $\psi_{s,t_0}(t)$  is the basic complex function, $s$ is the time scale defining the width of the wavelet, the symbol ``*''  denotes complex conjugation. Note that the time scales $s$ of the CWT allow a transition to the classical frequencies $f$ of the Fourier spectrum. Therefore, for convenience and simplicity of the interpretation of the results, we consider the results in the traditional plane $(f, t_0)$. 

The basic Morlet function $\psi_{s,t_0}(t)$  is often used to analyze the activity of the brain~\cite{aouinet2014electrocardiogram,ovchinnikov2011method}. By analogy with the Fourier power spectrum $F(f)$ for $f$-frequencies 
we can estimate the instantaneous frequency distribution of the CWT-energy
\begin{equation}
\label{kor31e} E(f, t_0)= |W(f, t_0)|^2.
\end{equation}

CWT allows to analyse an experimental EEG signal $x(t)$ simultaneously in the frequency and time domains. Moreover, the CWT excess property is well known, which is used to search for the fine time-frequency structure of experimental time series, which can be only poorly detected by conventional spectral methods~\cite{montefusco2014wavelets,maksimenko2018human,maksimenko2018multiscale,hramov2015wavelets}.

In this paper to analyze the oscillatory structure of the EEG, we use the CWT  ``skeleton'' method~\cite{cavalier2017maximum,runnova2017study,grubov2017development,grubov2017perception,hramov2017classifying}, which is based on a simplification of the entire surface $(f; t)$ of the experimental signal $x(t)$ by estimation only the main part of oscillatory activity.
For each moment of time $t_0$, it is necessary to search for the maxima in the instantaneous CWT spectrum $E(f, t)$~(\ref{kor31e}):
\begin{equation} \label{sceleton_dly_4ainika}
\forall f \in \delta f_{max}, E(f_{max}, t_0) \geq E(f, t_0),
\end{equation}
where $f_{max} = sc^{1}$ is the CWT skeleton, $\delta f_{max}$ is a $\delta$\,-\,region of the $f_{max}$ frequency component.
The first skeleton $f_1$ is the global maximum in the surface $W(f,t_0)$, i.~e. the value of the frequency in the signal spectrum, which correlates with the maximum of the oscillational energy. Further, for each moment of time, we detect a local maxima $sc^{1} > sc^{2} > \dots > sc^{j} > \dots >sc^{n_p}$, where $n_p$ is  the amount of skeletons.

Further, we limit the consideration of the oscillatory structure in EEG signals to the frequency band $\Delta f [1; 2,5]$~Hz. The choice of this ``slow'' frequency interval is due to the evidence in the literature on the presence of biomarkers in brain slow activity reflecting the BBB permeability changes~\cite{pavlov2020extended,Tian2020}. Thus, at each time moment $t_0$ we estimate which skeleton's number $N_0$ falls into a certain frequency range $\Delta f$. And next we calculate the average number of $N(t)$ skeletons falling into the band $\Delta f$ over the time interval $\Delta t = 50$~sec.

\begin{figure*}
\begin{center}
	\resizebox{0.75\linewidth}{!}{%
  	\includegraphics{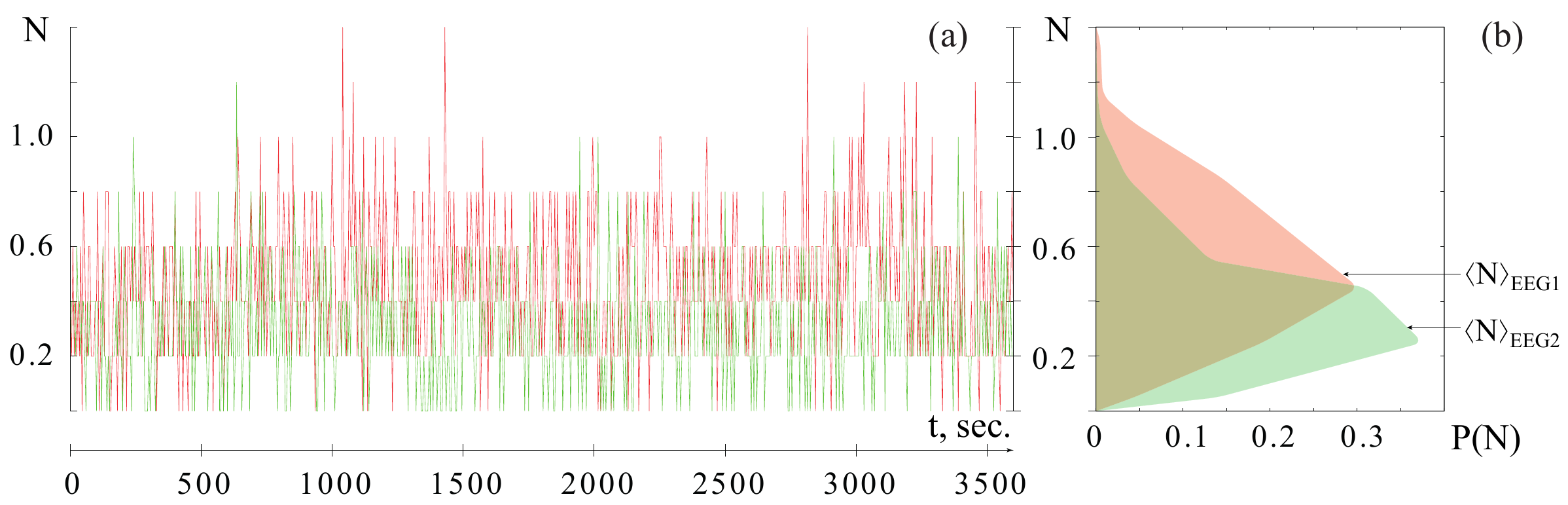} }
\end{center}
\caption{Results of evaluating the wavelet-based characteristics for animal \#\,3. Panel (a) shows the time-dependence of the $N$ average number of CWT-skeletons in the frequency range $\Delta f_1 [1; 2,5]$~Hz. The red line is the calculation result for EEG1, the green line corresponds to the characteristic for EEG2. The panel (b) represents the probability distributions of $N$ skeletons. Red color shows the distribution for EEG1 (no influence on the BBB-permeability), and green -- for EEG2 (after auditory impact). The arrows indicate the mean values for each distribution. }
\label{fig_wvlt}       
\end{figure*}

In Fig.~\ref{fig_wvlt}(a) the results of estimating the number of $N(t)$ skeletons for one experimental animal are presented. We observe that the number of patterns in the low frequency range $\Delta f_1$ increases after a powerful auditory impact~(EEG2). After evaluating the distribution of the number of patterns for records EEG1 and EEG2, their mean values differ $\langle N\rangle _{EEG1}\neq\langle N\rangle _{EEG2} $ (see Fig.~\ref{fig_wvlt}(b)). A similar situation is observed in the slow oscillatory activity of the EEG for all experimental animals.
Next, we use the detectable value $\langle N\rangle _{EEG2}$ as the threshold value that characterizes the moment of BBB permeability changes.

For each experimental animal, individual threshold values were obtained $\langle N\rangle _{EEG2} = \{0,31863;$ $0,45614;$ $0,462575;$ $0,485995;$ $0,52011\}$. Thus, we estimated the EEG-dynamics of the brain activity based on the  $N(t)$-dynamic. Generally, the value of dependence $N(t)$ in EEG2-record exceeded the threshold $\langle N\rangle_{EEG2}$ throughout first third of the recording. Next, this value $N(t)$ irregularly reduced towards the end of the EEG2-record, when, apparently, the influence of the sound impact was compensated by the neurophysiological system, and the BBB-permeability returned to its usual value. Based on this, we assumed that the excess of the $N(t)$ patterns number   of the  $\langle N\rangle _{EEG2}$  threshold value correlated with an increase in the BBB permeability. So, when the corresponding threshold value $N(t) \geq \langle N\rangle _{EEG2}$ is exceeded, we assumed that the permeability of the BBB was increased.

\section{ANN based method}
\label{sec:BBB_ANN}

\begin{figure*}
\begin{center}
	\resizebox{0.75\linewidth}{!}{%
  	\includegraphics{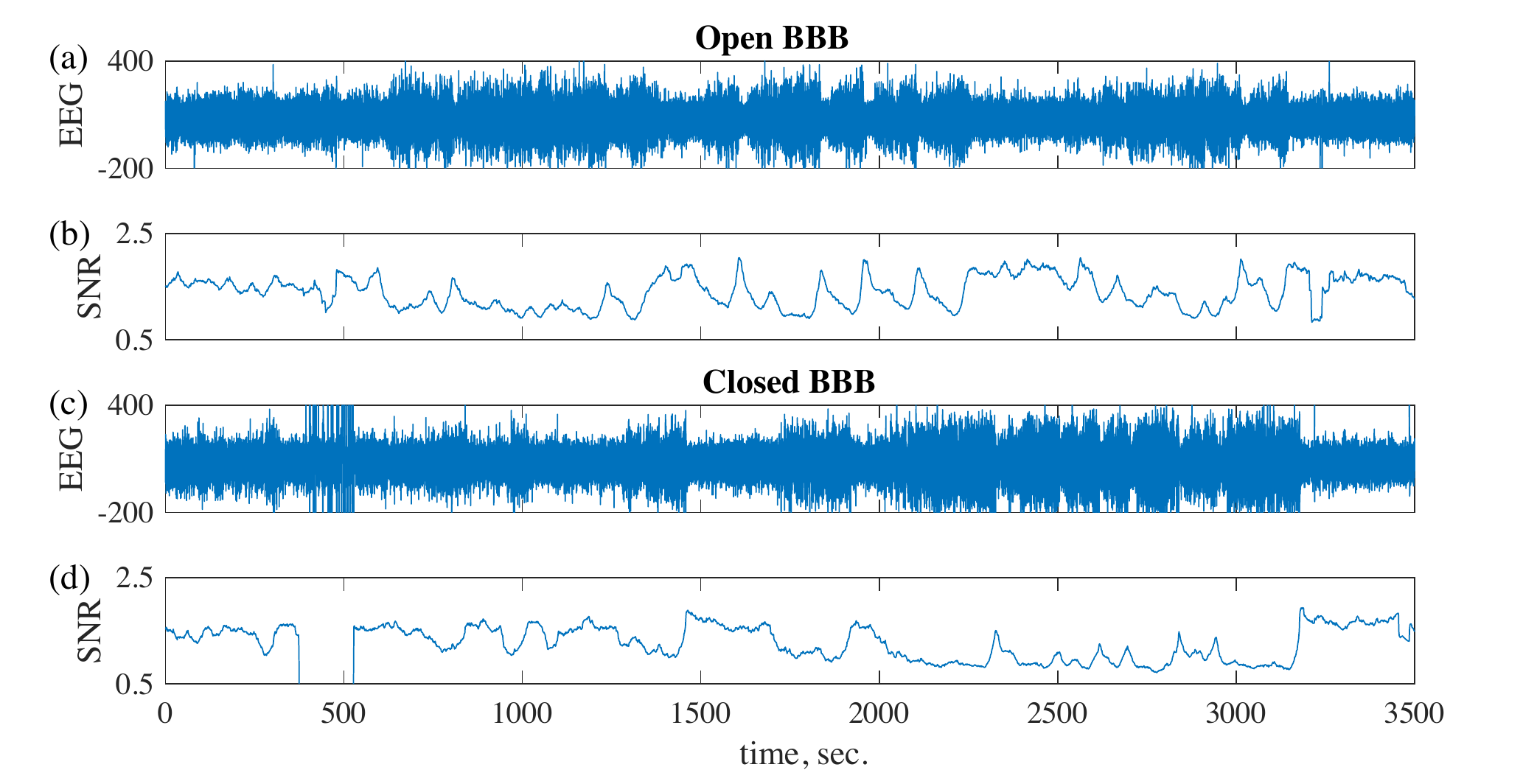} }
\end{center}
\caption{Examples of EEG signals (a,c) for animal \#\,1 and corresponding SNR characteristics (eq.~\ref{eq:snr}) shown in panels (b,d).} 
\label{fig:ann1}      
\end{figure*}

In order to estimate the permeability of the BBB, a deep neural network is used. The neural network was trained using an open-source library in Python \cite{Keras}. The use of EEG signals in their pure form is a non-trivial task for the ANN, since it requires a large number of neurons and layers. This dimension problem can be solved by specially prepared statistical characteristics calculated from these EEGs. There is a number of papers on sleep recognition, in which the EEG signals were not used by ANN, but their averages, variations and deviations from the mean \cite{Ronzhina2012,Hsu2013,Zhu2014,Hassan2015,Hassan2017}. In the same manner here we use a signal-to-noise ratio (SNR) as such a statistical characteristic. It is the ratio between the mean of the signal $\mu(\cdot)$ and its standard deviation $\sigma(\cdot)$: 
\begin{equation}
{SNR}(x)=\frac{\mu(x)}{\sigma(x)}.
\label{eq:snr}
\end{equation}
Each point of this characteristic is prepared for some averaging window. The window size is fixed to 30 seconds. In order to make the SNR realization smoother, the calculations were performed with a slowly shifting time window with a 1-second step. This leads to a new temporal determination of SNR with a time step of 1 sec, where each point corresponds to the SNR value obtained from a thirty-second window of EEG data. Figure \ref{fig:ann1} shows the EEG of one of the animals (panels a, c) and the calculated SNR characteristics (panels b, d).

To get the training SNR examples, we used EEG signals before (EEG1) and after (EEG2) artificially increased permeability of the BBB~\cite{SGl2018}. 
During the training, the fragments of SNR for the rat after music stimulation were marked as the expected response ``1'' (i.e. BBB is open), while fragments of rats before this stimulation corresponded to the response ``0'' (i.e. BBB is closed). Thus, the ANN response to an unknown signal can be interpreted as a degree of the averaged ``permeability'' of BBB over the considered period of time.

\begin{figure*}
\begin{center}
	\resizebox{0.75\linewidth}{!}{%
  	\includegraphics{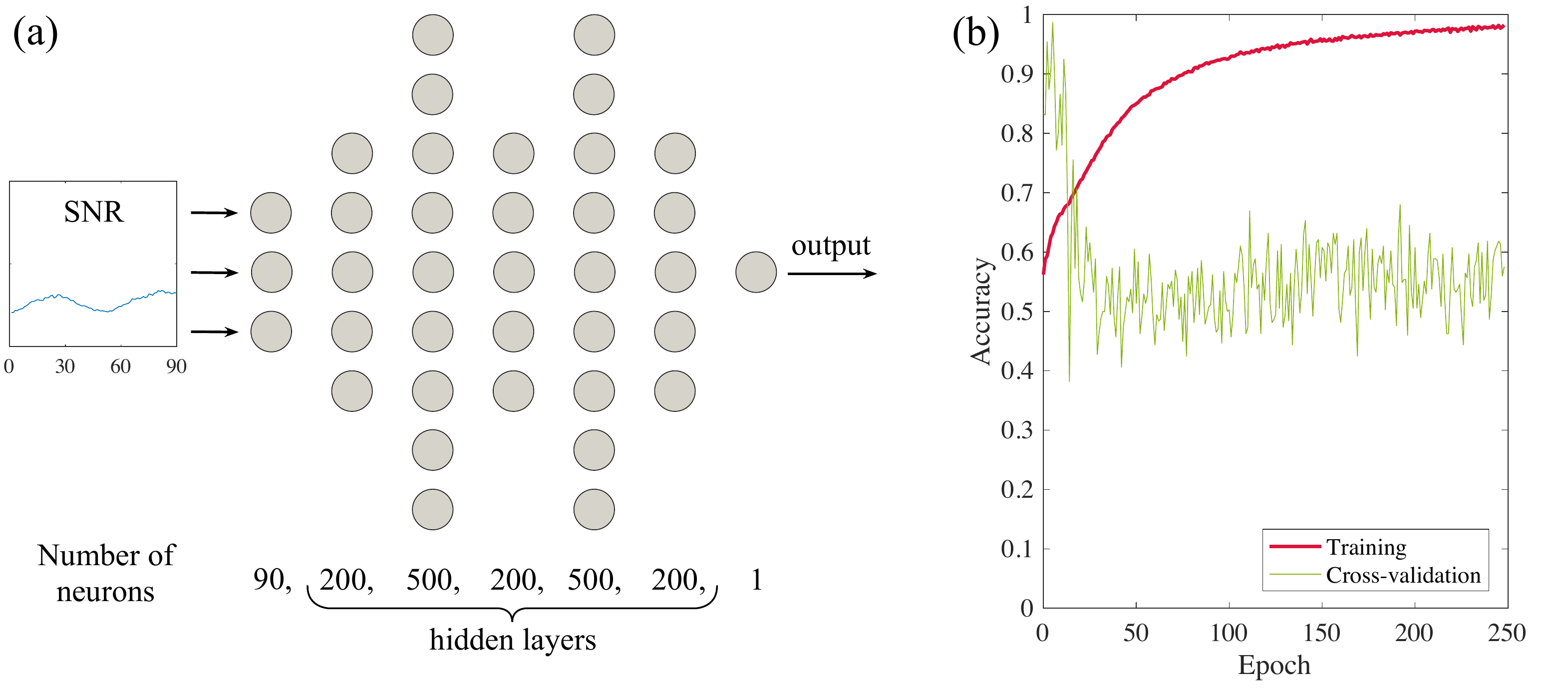} }
\end{center}
\caption{Schematic diagram of considered DNN (panel a) and the dynamics of accuracy during training depending on the epoch number (panel b)}
\label{fig:ann2}       
\end{figure*}

The input layer of our ANN consists of 90 neurons. Each time the network receives a 90-second SNR realization as an input. A smaller number of input neurons leads to slower training and a sharp jump into overtraining. A larger number leads to inaccuracies in the temporal marking, since it gives the response delay of more than 1.5 minutes. Also a few DNN configurations with constant numbers of neurons in the hidden layers 200, 500, 1000 have been considered. A network with a variable number of neurons in hidden layers (500$\times$200$\times$500$\times$200) has shown the optimal results during training and testing. The network scheme and the training process are shown in Fig.~\ref{fig:ann2}. All the layers have the sigmoid activation function $f(x)=1/(1-e^{-x})$.

\begin{figure*}
\begin{center}
	\resizebox{0.75\linewidth}{!}{%
  	\includegraphics{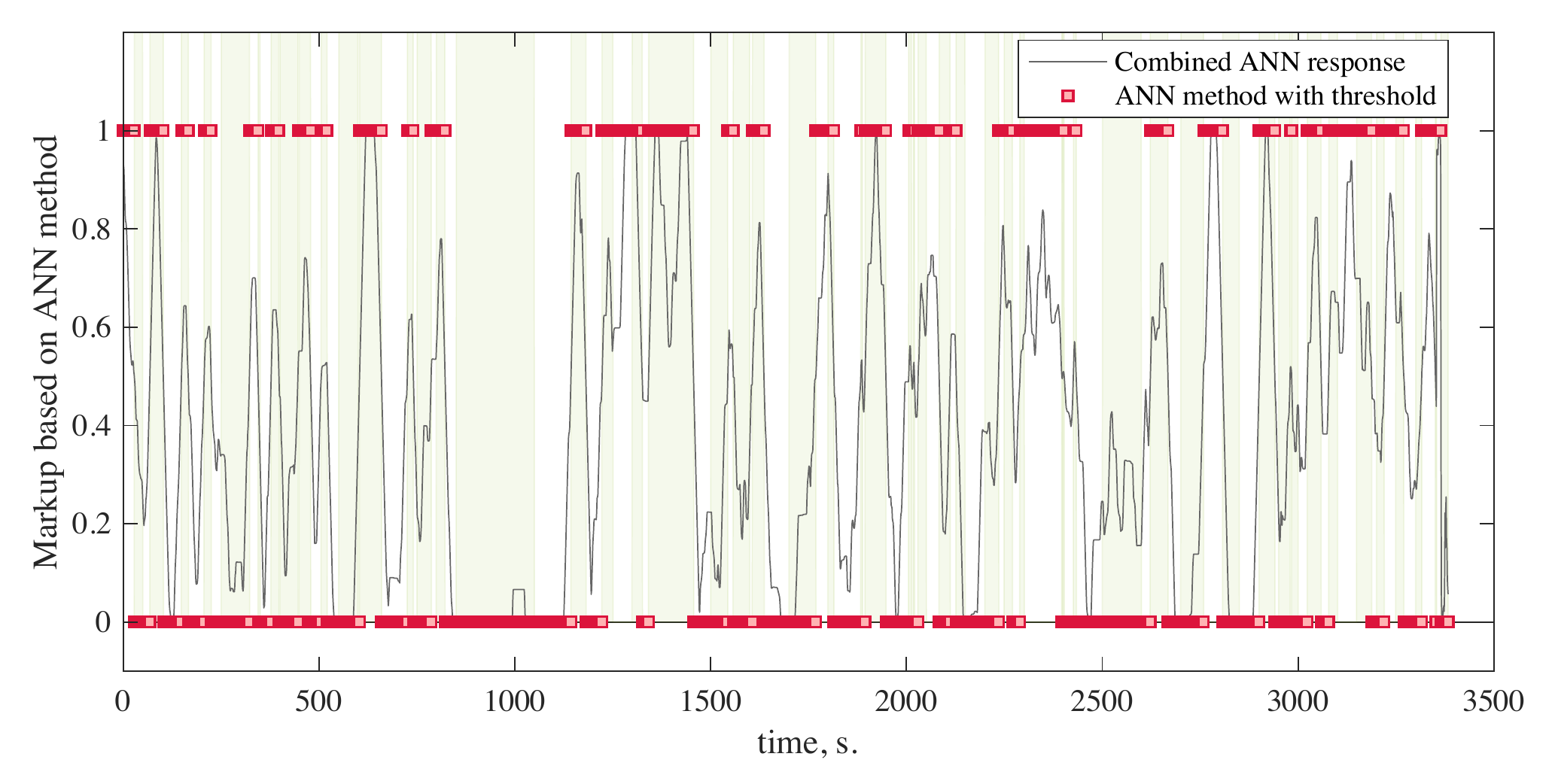} }
\end{center}
\caption{Temporal evolution of ANN responses for animal \#\,1 after music-induced OBBB. The multiplication of network responses for both EEG channels is shown in gray. Red points correspond to the binary classification after introducing the threshold 0.5. Green areas show time intervals in which both markings coincide: (1) based on ANN answers and (2) wavelet method from Sect.~\ref{sec:BBB_vawelet}.}
\label{fig:ann3}       
\end{figure*}

For each of the rats the ANN was trained on the data of the remaining rats. This increased the number of training and testing implementations. Each EEG recording was done using two channels. The SNR and ANN markup were prepared for both signals separately. To increase accuracy, the network responses for both channels were multiplied. Thus, if the network recognises the inputs from both channels are recognized simultaneously as ``1'' (BBB is open), then the final answer is ``1''. If one channel leads to the answer ``0'' and the second one leads to the answer ``1'', then the final response is ``0''. Since the network produces a real number from 0 to 1, the multiplication of responses from two channels can lead to an increase in intermediate values. To eliminate this, the threshold 0.5 has been introduced. Since then, the answers $y \geq 0.5$ are regarded as ``1'' (BBB is open), while responses $y < 0.5$ complies with ``0'' (BBB is closed). Figure \ref{fig:ann3} shows a temporal evolution of ANN responses for an animal with an artificially open barrier, whose signals were shown in Fig.~\ref{fig:ann1}(a,b). The gray line represents the result of multiplying the signals from both channels, and red pints show the final answer with the threshold 0.5. 

\begin{figure*}
\begin{center}
	\resizebox{0.75\linewidth}{!}{%
  	\includegraphics{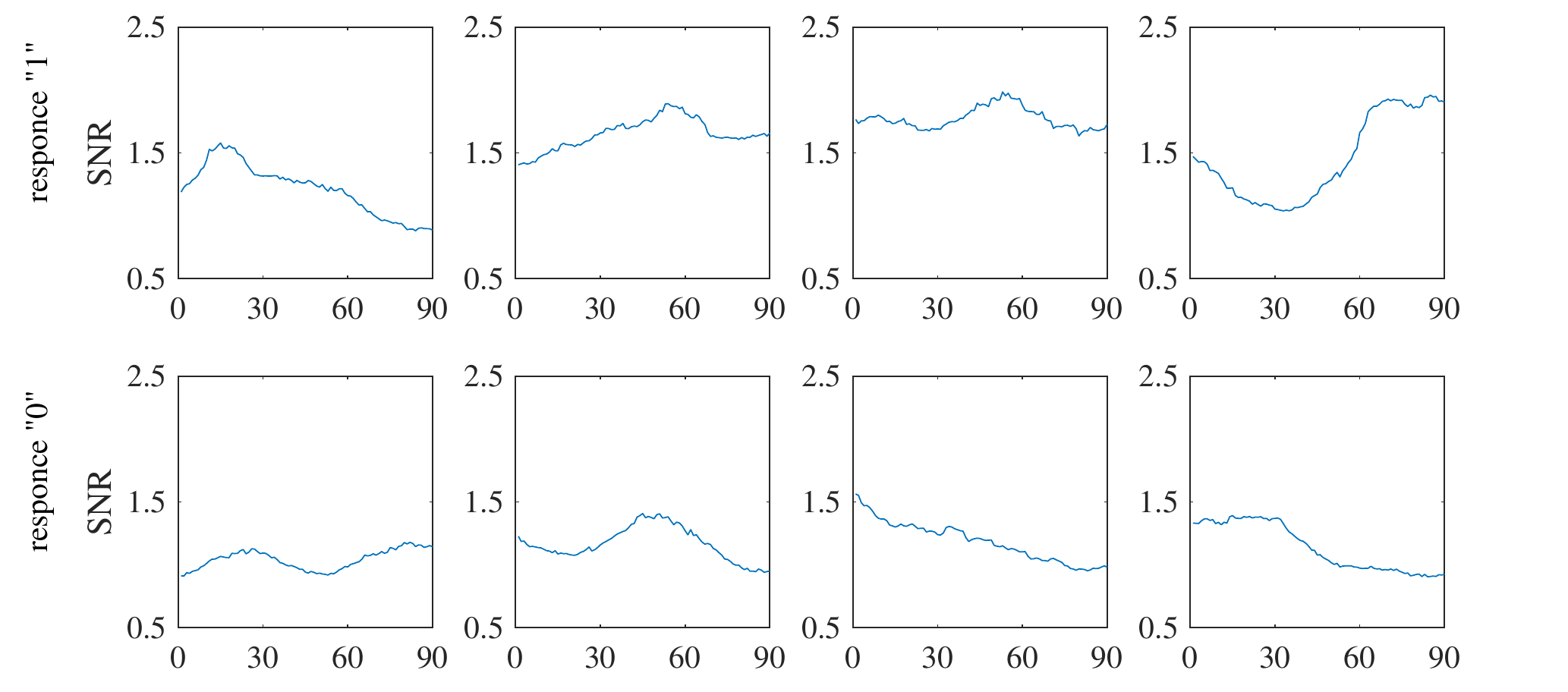} }
\end{center}
\caption{Examples of input SNR implementations on which the neural network exhibited a similarity to OBBB (top panels) and closed barrier (bottom panels).}
\label{fig:ann4}       
\end{figure*}

The ANN found 32\% of the data similar to the open barrier for implementations after music-induced OBBB. In the case of free behaviour, this percentage was 24\%. These ratios are averaged over all considered animals. In order to exclude the peculiarities of the training, the networks have been trained 5 times for each animal with an accuracy of at least 96\% (approximately 250 epochs). Table~\ref{tab:ann1} shows the percentage of fragments recognized as OBBB for one animal after 5 trainings. Afterward, the general percentage of fragments, recognized as OBBB, has been averaged over all animals and over all five trainings.  Figure \ref{fig:ann4} shows examples of SNR obtained using fragments of EEG signals recognized as signals with open BBB (top panels) and closed BBB (bottom panels). 

\section{Comparison of two independent methods of data processing}

Figure \ref{fig:ann3} shows the intersection of both methods from Sect.~\ref{sec:BBB_vawelet} and \ref{sec:BBB_ANN} by green background color. We find that both methods are in a good agreement with each other. To test this numerically, the so-called F-measure \cite{fscore} is applied, which is often used for statistical analysis of binary classification. It is the harmonic mean between precision and recall. The precision is the number of correctly identified positive results divided by the number of all positive results, including those not identified correctly, and the recall is the number of correctly identified positive results divided by the number of all samples that should have been identified as positive. For such an assessment, one can choose one of the methods as true, for example, the wavelet method, and then count the number of matches of both methods for the answer ``yes'' (true-positive, TP), the number of matches for the answer ``no'' (true-negative, TN), as well as mismatches (false-positive, FP, and false-negative, FN). Based on this, the F-measure is calculated as follows:
\begin{equation}
\label{eq:ann1}
P=\frac{TP}{TP+FP}, R=\frac{TP}{TP+FN}, F^+=2 \frac{PR}{P+R}
\end{equation}

Thus, the F-measure $F^+$ is an evaluation of the correctness for the answers ``yes''. If we replace all letters ``P'' by ``N'' in Eq.~(\ref{eq:ann1}), and vice versa, then this will be an estimate of the correctness for the answer ``no'', which we denote as $F^-$. Table~\ref{tab:ann1} shows the results of calculating the F-measure for five ANN trainings, as well as the averaged characteristics. Thus, according to the calculations of the F-measure, we can conclude that the intersection of 35~\% in the answers that the barrier is open, and 75~\% in the answers about the closed barrier. These values, averaged over all animals, are 28~\% and 70~\%.

\begin{table}
\centering
\label{tab:ann1}
\begin{tabular}{lllllll}
\hline\noalign{\smallskip}
Training: & Tr.1 & Tr.2 & Tr.3 & Tr.4 & Tr.5 & \textbf{Averaged}  \\
\noalign{\smallskip}\hline\noalign{\smallskip}
\% of open barrier in realization with an open barrier	 & 35 & 22 & 21 & 38 & 27 & \textbf{28.5} \\
\% of open barrier in realization with free behaviour & 24 & 18 & 13 & 24 & 20 & \textbf{19.8}\\
F-measure on responses ``BBB is open'', $F^+$ & 0.38 & 0.35 & 0.32 & 0.38 & 0.34 & \textbf{0.35}\\
F-measure on responses ``BBB is closed'', $F^-$ & 0.73 & 0.77 & 0.79 & 0.72 & 0.75 & \textbf{0.75} \\
\noalign{\smallskip}\hline
\end{tabular}
\caption{Percentage of fragments recognized as OBBB and not, and F-measures of animal \#\,1 for five ANN trainings on fragments of the other animals.}
\end{table}

\begin{figure*}
\begin{center}
	\resizebox{0.75\linewidth}{!}{%
  	\includegraphics{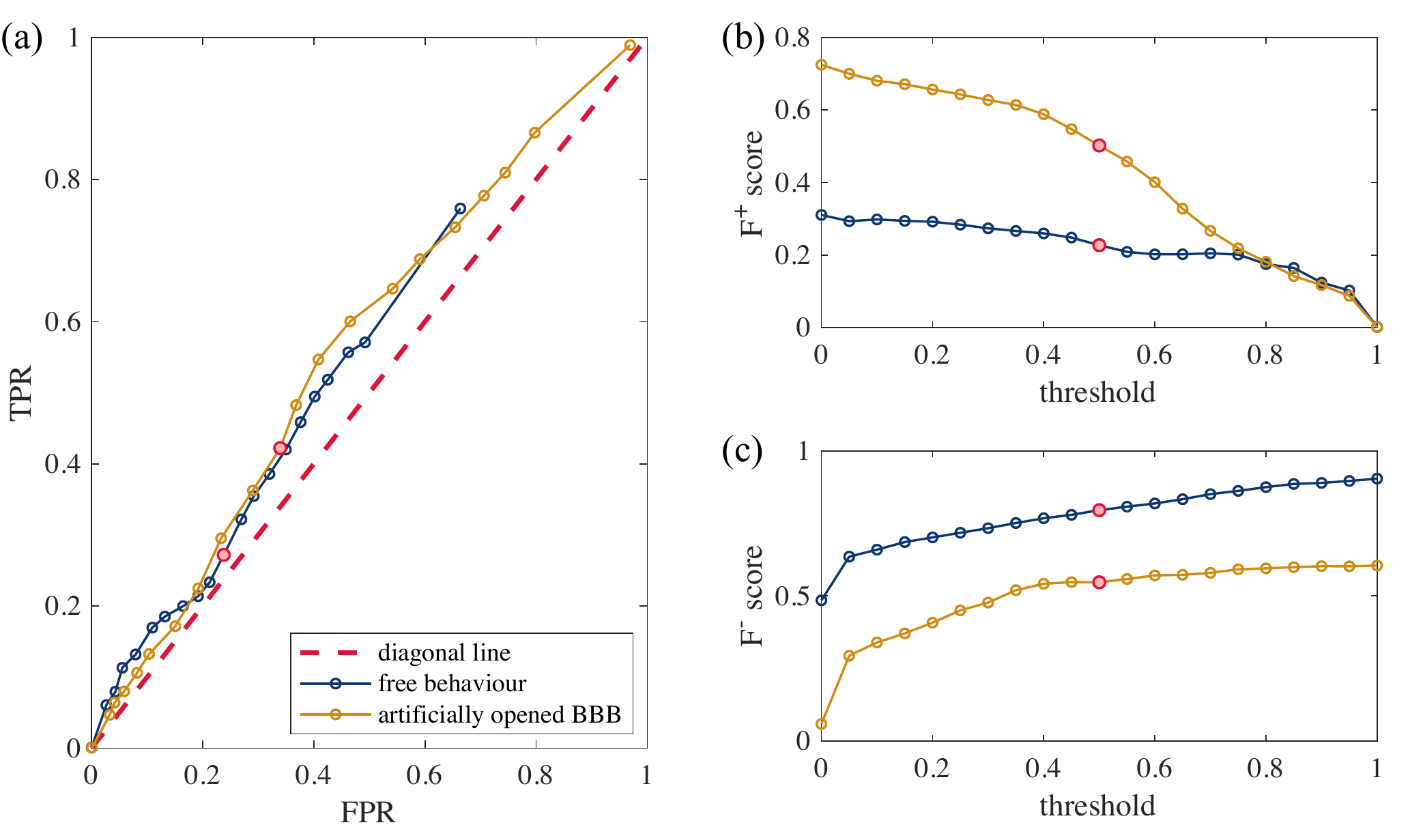} }
\end{center}
\caption{Statistical comparison of both methods using ROC-curve (panel a), and F-scores (eq.~\ref{eq:ann1}) for positive (b) and negative (c) responses.}
\label{fig:ann5}       
\end{figure*}

Another measure that allows one to assess the quality of the classification is the receiver operating characteristic (ROC-curve). The ROC curve is created by plotting the true positive rate ($TPR=\frac{TP}{TP+FN}$) against the false positive rate ($FPR=\frac{FP}{FP+TN}$) at various threshold settings \cite{Fawcett2006,Powers2011}. The classifier is considered to be good if the ROC-curve is above the diagonal (see Fig.~\ref{fig:ann5}). However, the degree of success of the classifier is roughly estimated by how far this curve is above the diagonal, and, accordingly, how large is the area below it. Figure \ref{fig:ann5} shows the ROC curves for different threshold values in cases of artificially opened BBB (orange color) and free behaviour (blue color), when the rat can either sleep or be awake. The markings obtained by the method, described in Sect.~\ref{sec:BBB_vawelet}, were used as a verification data. Both curves are above the diagonal, which indicates the similarity of the methods. In addition, the figure shows the results of calculating the F-measures for both implementations with variation of the threshold parameter. The points corresponding to the $0.5$ threshold are highlighted by red. As follows from the graphs of F-measures, the threshold $0.5$ is optimal for positive and negative answers. An increase in the threshold leads to an increase in false positives, and a decrease leads to a loss of sensitivity when the network says that the barrier is always closed.

\section{Conclusions}

This work is devoted to the investigation of sustained changes in brain activity after music-induced OBBB. BBB permeability is analyzed using two different methods -- wavelet estimation of the oscillatory characteristics of the EEG and machine learning using SNR characteristics. The wavelet analysis demonstrated an increase in the number of arising oscillations in the low-frequency region (1 -- 2.5 Hz), which is universal for all animals. However, the threshold for the increase in oscillatory activity in these frequencies is highly individual for each animal.

The second method is based on the search of special features of SNR realisation which are common for animals with OBBB. The use of a machine learning approach allows us to demonstrate a high level of universality of the revealed ``portrait'' of the electrical activity of the brain after an artificial increase in the permeability of the BBB. 

Despite the fact that both methods are based on different characteristics, they show a good overlap both for animals with artificially open BBB and for animals in free behaviour without any influences. Quantification of the methods agreement was based on ROC curves and F-measures of positive and negative responses. Both methods recognized a non-zero probability of opening the barrier for animals with normal behaviour. It is possible that this effect is related to the constancy of brain activity patterns during the physiological state of sleep. A number of authors have linked spontaneous changes in BBB permeability to certain stages of sleep, promoting a repair and clearance of brain tissue \cite{semyachkina2017application,zinchenko2019pilot,pavlov2020extended}.

In the future, we plan to increase of the experimental groups and to include instrumental methods of optical control of the BBB permeability. The preliminary analysis of brain activity during the sleeping state of the animals is of our interest. Probably, some sleep stages would be accompanied with the brain activity, similar to the results of music-induced OBBB.

\section{Acknowledgements}
This work has been supported by the RF Government Grant No. 075-15-2019-1885 in part of the biological interpretation. In the part of the development of numeric method of data analysis this work has been supported by the Council for Grants of the President of the Russian Federation for the State Support of Young Russian Scientists (project no. MD-645.2020.9).

%

\end{document}